\documentclass[aps,preprint,amsmath,amssymb,showpacs,amsfonts]{revtex4}
\usepackage{epsfig}
\usepackage{graphicx}
\usepackage{dcolumn}
\usepackage{bm}

\newcommand{\beq}{\begin{equation}}
\newcommand{\eeq}{\end{equation}}
\newcommand{\bea}{\begin{eqnarray}}
\newcommand{\eea}{\end{eqnarray}}

\begin{document}

\title{The Penrose Inequality and the Fluid/Gravity Correspondence}

\author{Yaron Oz}
\author{Michael Rabinovich}
\affiliation{Raymond and Beverly Sackler School of
Physics and Astronomy, Tel-Aviv University, Tel-Aviv 69978, Israel}

\date{\today}

\begin{abstract}
Motivated by  the fluid/gravity correspondence, we consider the Penrose inequality in the framework of
fluid dynamics.
In general relativity, the Penrose inequality relates the mass and the entropy associated with a gravitational background.
If the inequality is violated by some Cauchy data, it suggests a creation of a naked singularity, thus providing means to
consider the cosmic censorship hypothesis.
The analogous inequality in the context of fluid dynamics can provide a valuable tool in the study of
finite-time blowups in hydrodynamics.
We derive the inequality for relativistic and nonrelativistic fluid flows in general dimension.
We show that the inequality is always satisfied at the ideal fluid order.
At the leading viscous order, the inequality may be violated by relativistic
fluid flows, while it is always satisfied by nonrelativistic incompressible flows.
The inequality may be violated at the next to leading viscous order by both relativistic and nonrelativistic flows.

\end{abstract}
\maketitle

\tableofcontents

\section{Introduction}

Compressible hydrodynamics allows singularities in the form of shock waves, that is a jump in the velocity vector field.
Incompressible hydrodynamics does not exhibit shock waves, and by singularities one means blowups of the derivatives of the velocity vector field.
It is often expected that the viscosity terms act as regulators that prevent the blowups.

The basic question concerning singularities in the hydrodynamic description, is whether starting with appropriate initial conditions, where the velocity vector field and its derivatives are bounded, can the system evolve such that it will exhibit within a finite time a blowup of the derivatives of the vector field.
Mathematically, this important question has been presented as one of the
 millennium problems posed by the Clay Mathematics Institute \cite{clay}.
Physically, such singularities if present, indicate a breakdown of the effective hydrodynamic description at long distances and imply
that some new degrees of freedom are required.

The issue of hydrodynamic singularities has an analogue in gravity. Given an appropriate Cauchy data, will the evolving space-time
geometry exhibit a naked singularity, i.e. a blowup of curvature invariants  and the energy density of matter fields at a point not covered by a horizon.
The cosmic censorship conjecture in general relativity (for a review see \cite{Wald:1997wa}), states in its weak form that any space-time singularity is generically enclosed by an event horizon, that is
a region that no light rays can escape from. Mathematically, naked singularities can exist, but physical considerations of causality seem to require that it should not be visible to distant observers. In its strong form, the conjecture states that even inside the black hole a falling observer will never "see" the singularity, i.e.
time-like singularities never occur generically.

The Penrose inequality \cite{Penrose:1973um} (for a review see e.g. \cite{Mars:2009cj}) is a conjecture relating the mass and the horizon area of any Cauchy initial data that if violated leads to a
space-time naked
singularity. Let us briefly review Penrose's argument in asymptotically flat four-dimensional space-time. Consider a Cauchy data $({\cal M}^3,g,K)$, where ${\cal M}^3$ is a 3-manifold with a complete Riemannian metric $g$ and an extrinsic curvature $K$, satisfying
the Gauss-Codazzi equations.
In particular, let an asymptotically flat initial Cauchy data with a mass $M_0$ and event horizon area $A_{H0}$  evolve with time.
 The solution to Einstein equations with this initial data is expected to settle down at long time to a Kerr black hole solution with mass $M$ and horizon area $A_H$, where the inequality (we use $c=G_N=1$)
\bea
M\geq \sqrt{A_H/{16 \pi}}
\label{kerr}
\eea
 holds. The inequality is saturated by the Schwarzshcild black hole.

 By the Hawking area theorem \cite{Hawking:1971tu}, the event horizon area does not decrease with time $A_H \geq A_{H0}$.
 The mass, on the other hand, cannot increase and may only decrease due to radiation loss $M \leq M_0$. Thus, we get the chain of inequalities
 \begin{equation}
 M_0\geq M \geq \sqrt{A_H/{16 \pi}}\geq \sqrt{A_{H0}/{16 \pi}} \ ,
\end{equation}
and the initial Cauchy data should also satisfy the Penrose inequality $ M_0\geq \sqrt{A_{H0}/{16 \pi}}$.

The argument relies on the Hawking area theorem and the relaxation at late times to a Kerr solution, both assume the weak censorship hypothesis.
Therefore, finding Cauchy data
that violates the Penrose inequality implies finding a solution to Einstein equations in which a naked singularity is created.
Proving that the inequality is satisfied does not prove the cosmic censorship conjecture but provides an important support for it.
Note, that there are other less likely ways to interpret a violation of the inequality, e.g. that the system never reaches equilibrium.

In fact, we can formulate an analog of the Penrose inequality for any dynamical system. Consider a system, where the total energy does not increase. We replace the area increase theorem by the second law of thermodynamics, that is the total entropy never decreases. The system is assumed to relax at late time to a thermodynamical equilibrium state.
The inequality relation between the mass and the entropy is
simply an equilibrium relation, while by repeating Penrose's argument, the inequality should hold away from equilibrium.

The fluid/gravity correspondence relates in a derivative expansion two non-linear systems, the fluid dynamics described by the Navier-Stokes equations
and gravity described by general relativity.
In one particular (conformal) setup of the correspondence \cite{Bhattacharyya:2008mz}, the Navier-Stokes equations in $d$ space-time dimensions are identified with the constraint Einstein equations of a deformed black brane gravitational background in one higher dimension.
Alternatively, they are identified with the Gauss-Codazzi equations governing the evolution of the black brane event horizon \cite{Damour,Eling:2009pb,Eling:2009sj}.
Motivated by  the fluid/gravity correspondence, we consider the Penrose inequality in the framework of
fluid dynamics, which can provide a valuable tool in the study of
finite time blowups in hydrodynamics.
We derive the inequality for relativistic and nonrelativistic fluid flows in general dimension
and study its implications.

The paper is organized as follows.
In section II we briefly review the fluid/gravity correspondence in the conformal case.
We then derive the inequality for relativistic ideal hydrodynamics, where we consider both the conformal
and non-conformal cases. We show that the inequality is always satisfied at the ideal fluid order, and present
a class of solutions that develop a shock wave singularity, while still satisfying the inequality.
This is curious, though of course there is no inconsistency, since a non-violation of the inequality is only a sufficient condition for not having a
finite-time blowup and not a necessary one.
In section III we derive the inequality at the first and second viscous order for relativistic fluids. The inequality is nontrivial
and maybe be violated.
In section IV we show that the inequality is always satisfied by nonrelativistic incompressible fluids. However, this is
not the case when considering subleading corrections to the incompressible Navier-Stokes equations.
Section V is devoted to a discussion.

\section{The Penrose inequality in hydrodynamics}

In this section we will formulate the Penrose inequality for fluid dynamics. We will use the fluid/gravity correspondence to motivate
the inequality. However, the actual derivation does not require the use of the correspondence.

\subsection{The fluid/gravity correspondence: conformal case}

Consider the vacuum $(d+1)$-dimensional Einstein equations with a negative cosmological constant
\bea
R_{MN}+d g_{MN}=0,~~~~~~~ R=-d(d+1)~~~~~(M,N=0,..,d) \ .
\label{EQ}
\eea
We will use the notation $X^M = (x^{\mu}, \mu=0,...,d-1, x^d=r)$.
One solution of these equations is the boosted  black brane with $AdS_{d+1}$ asymptotics
\bea ds_0^2=-2u_\mu dx^\mu dr+\frac{1}{b^d r^{d-2}}u_\mu u_\nu dx^\mu dx^\nu +r^2\eta _{\mu \nu}dx^\mu dx^\nu \ ,
\label{zerometric} \eea
with the $d$-dimensional velocity vector $u^{\mu} = (\gamma,\gamma v_i),u_\mu u^\mu =-1$, $b$ related to the Hawking temperature by
\bea b=\frac{d}{4 \pi T} \ ,
\label{b} \eea
and $\eta_{\mu\nu}= diag(-1,1..,1)$ is the $d$-dimensional Minkowski metric.

The metric (\ref{zerometric}) is an exact solution of the equations (\ref{EQ}) when $b$ and $u_{\mu}$ are constant.
This is the thermal equilibrium state.
Next, we get out of equilibrium by deforming the solution and allowing $b(x^{\mu}), u_{\mu}(x^{\mu})$.
With these, the metric (\ref{zerometric}) is no longer a solution of the equations (\ref{EQ}).

One can, however, correct the metric (\ref{zerometric}) by adding derivatives of $b(x^{\mu}), u_{\mu}(x^{\mu})$.
We will have in general
\begin{equation}
ds^2 = ds_0^2 + ds_1^2 + ds_2^2...
\label{deformed}
\end{equation}
where the lower index of $ds^2$ denotes the order of the derivative terms that is contains and $ds_0^2$ is the metric (\ref{zerometric}).
Requiring the deformed metric (\ref{deformed}) to be a solution of (\ref{EQ}) up to the neglected order in derivatives,
is equivalent the order by order derivative expansion of conformal hydrodynamics \cite{Bhattacharyya:2008mz}.
More specifically, the set of $d+1$ constraint Einstein equations can be written as the conservation law of a conformal hydrodynamic
stress energy tensor in flat $d$-dimensional space with specific transport coefficients
\bea \partial_\mu T^{\mu \nu}=0  \ . \eea
Thus, the gravity solution provides a dual description in terms of fluid dynamics.
Equivalently, one arrives at the fluid/gravity correspondence starting from the AdS/CFT correspondence, and considering
the hydrodynamics of the CFT and its gravitational dual description.
Alternatively, the hydrodynamic equations are identified with the Gauss-Codazzi equations governing the evolution of the black brane null event horizon \cite{Damour,Eling:2009pb,Eling:2009sj}.
The various schemes are related by an RG flow in the radial direction
\cite{Bredberg:2010ky}.

The hydrodynamic stress energy tensor $T_{\mu\nu}$ can also be calculated at the $d$-dimensional boundary \cite{Balasubramanian:1999re}.
We write the $(d+1)$-dimensional metric in an ADM-like decomposition
\bea
ds^2=N^2 dr^2+\gamma_{\mu \nu}(dx^\mu+N^\mu dr)(dx^\nu+N^\nu dr) \ ,
\eea
where $\gamma_{\nu \mu}$ is the boundary ($r=const.$) metric, $N$ and $N^\mu$ are the lapse and shift functions.
The extrinsic curvature tensor of the boundary $\partial M$ is defined by
\bea
K_{\mu \nu}=-\frac{1}{2N}(\partial _r \gamma _{\mu \nu}- \nabla_\mu N_\nu- \nabla_\nu N_\mu) \ ,
\eea
and the boundary stress energy tensor reads
\bea T_{\mu \nu}=lim_{r \rightarrow \infty} [\frac{r^{d-2}}{64 \pi ^2}(K_{\mu \nu}-K \gamma _{\mu \nu}-(d-1) \gamma _{\mu \nu})] \ .
\label{boundarystress}
\eea

\subsection{The Penrose inequality in conformal hydrodynamics}

The hydrodynamic stress energy tensor of an ideal relativistic neutral fluid is
\bea
T_{\mu\nu} = (\epsilon + p)u_{\mu}u_{\nu} + p \eta_{\mu\nu} \ ,
\label{stressrel}
\eea
where $\epsilon$ is the energy density and $p$ is the pressure.
Conformal invariance implies that the stress energy tensor is traceless, thus we have the equation of state
$\epsilon = (d-1)p$. The only dimensionfull parameter is the temperature $T$ and we have that $\epsilon = (d-1)p \sim T^{d}$.
The boundary stress tensor (\ref{boundarystress}) corresponding to the stationary metric (\ref{zerometric}) is
that of ideal conformal neutral fluid and reads \cite{Bhattacharyya:2008jc,Haack:2008cp}
\bea
T_{\mu\nu}=\frac{1}{16 \pi b^d} (\eta_{\mu \nu}+du_\mu u_\nu) \ .
\eea

Let us derive the analog of the Penrose inequality in this case. To this end we need to define the energy (mass), the horizon area and
the equilibrium relation.
The energy ($E=M$) is given by
\bea E=\int T_{00} d^{d-1}x =\frac{d \gamma^2-1}{16 \pi b^d} \int d^{d-1}x \label{mass0} \ , \eea
while the horizon area is defined via the entropy $S$.
Consider a conformal fluid at rest $(u_\mu=(1,0,0,0))$ satisfying the thermodynamic relation
\bea \epsilon+p= Ts  \ , \eea
where $s$ is entropy density, which in the conformal case reads
\bea s=\frac{\epsilon}{T}\frac{d}{d-1}=\frac{1}{4b^{d-1}}  \ , \eea
and the horizon area is
\bea A =4S =\int \frac{1}{b^{d-1}} d^{d-1}x  \ ,\eea
so we get that the equilibrium relation between energy and entropy
\bea E=\frac{d-1}{16 \pi}(4S)^{\frac{d}{d-1}} \ ,  \label{Peq} \eea
where we normalized the volume of space to one.
The fluid analog of the Penrose inequality is therefore
\bea E\geq\frac{d-1}{16 \pi}(4S)^{\frac{d}{d-1}}  \ . \label{PI} \eea
One way to verify the direction of the inequality is to consider the black brane with angular momentum.

Consider as an example the case when the fluid is not at rest.  The inequality is slightly generalized in this case
by replacing the entropy density $s$ by the
time component of entropy current $s^{\mu} = s u^{\mu}$ \cite{Bhattacharyya:2008xc}, that is $s^{0} = s \gamma$.
For any $\gamma\geq 1$ we have
\bea
\frac{d \gamma^2\ -1}{d-1}\geq \gamma^{\frac{d}{d-1}} \ .
\eea
Thus, by H\"{o}lder's inequality we get
\bea
\int \frac{d \gamma^2\ -1}{b^d(d-1)}\geq \int \frac{\gamma^{\frac{d}{d-1}}}{b^d} \geq (\int \frac{\gamma}{b^{d-1}})^\frac{d}{d-1} \ .
\label{holder}
\eea
The inequality (\ref{PI}) holds for any $\gamma \geq 1$, and is an equality only when $\gamma=1$, that is for a fluid at rest.

Note, that although motivated by the fluid/gravity correspondence the argument that leads to the inequality for fluids is rather general and
is independent of having a gravitational description.
It assumes that at late times the system relaxes to an equilibrium state, whose thermodynamic relation gives
the inequality relation. One then assumes that mass (energy) is a non-increasing, that is no energy is pumped into the system, and that
the entropy is non-decreasing by the second law of thermodynamics.
Thus, starting with initial data $(E_0,S_0)$ one has
\bea E_0\geq E \geq \frac{d-1}{16 \pi}(4S)^{\frac{d}{d-1}} \geq \frac{d-1}{16 \pi}(4S_0)^{\frac{d}{d-1}} \ . \label{PI1} \eea
The Penrose inequality is a condition to be imposed on any initial spacetime configuration (Cauchy data). When this condition fails, the evolution
 in time is expected to exhibit a naked singularity.
Likewise, The inequality (\ref{PI}) is a condition that we impose on any initial fluid data that satisfy the hydrodynamic equations. If the condition fails, we expect the fluid
to exhibit a finite-time blowup of the velocity field gradients.
Note, however, that a violation of the Penrose inequality is a sufficient condition for a creation of a naked singularity, but is not a necessary one.
Likewise, a violation of the hydrodynamic inequality is only a sufficient condition for a singularity.

It is quite important not to link directly singularities in the hydrodynamic description to
naked singularities in gravity. The reason being that the fluid/gravity correspondence is established order by order in a derivatives expansion.
An appearance of singularities signals the breakdown of a derivative expansion and for a direct link between the two descriptions
one needs to go beyond the established relation.

\subsection{Nonconformal fluids}

We can easily generalize the inequality to non-conformal fluids with a general equation of state.
As an example consider an equation of state
of the form $E=\sigma p$ where $\sigma \neq d-1$.
An example of a dual gravitational background for such an equation of state was constructed in \cite{Kanitscheider:2009as}.
A study of the universal statistics of such hydrodynamics in $d=2$ has been carried out in \cite{Liu:2010jg}.
By dimensional analysis $p=aT^d$ and
\bea
\epsilon=\sigma a T^d, \ \ s=\frac{\epsilon+p}{T}=(\sigma+1)aT^{d-1} \ .
\eea
The relation between the energy and entropy reads
\bea
E=\frac{\sigma}{(\sigma+1)^{\frac{d}{d-1}}a^\frac{1}{d-1}}S^\frac{d}{d-1} \ ,
\eea
and the inequality in this case is
\bea
E \geq \frac{\sigma}{(\sigma+1)^{\frac{d}{d-1}}a^\frac{1}{d-1}}S^\frac{d}{d-1} \ .
\label{NC}
\eea
The inequality (\ref{PI}) is a special case of (\ref{NC}) with $\sigma=d-1$ and $a = \frac{1}{16\pi}\left(\frac{4 \pi}{d}\right)^d$.

\subsection{Simple relativistic waves}

As stated above, a violation of the Penrose inequality is a sufficient condition for a creation of a naked singularity, but is not a necessary one.
Likewise, a violation of the hydrodynamic inequality is only a sufficient condition for a singularity and not a necessary one.
In this subsection we give an example of initial relativistic conformal ideal fluid flow configurations leading to a formation of a singularity, despite
satisfying the Penrose inequality.

The hydrodynamic equations for a relativistic conformal ideal fluid
can be written in the form
\bea
\partial_{\mu}u^{\mu} + (d-1) D ln T = 0,~~~~~~~~a_{\sigma} + P_{\sigma}^{\mu}\partial_{\mu} ln T = 0 \ ,
\label{idealeq}
\eea
where $D=u^{\mu}\partial_{\mu}$, $a_{\sigma}= u^{\mu}\partial_{\mu}u_{\sigma}$.
One obtains this form of the equations by the projection of the stress energy conservation equations as $u_{\nu}\partial_{\mu}T^{\mu\nu}$ and
$P^{\sigma}_{\nu}\partial_{\mu}T^{\mu\nu}$, with
\begin{equation}
P_{\mu\nu} = \eta_{\mu\nu} + u_{\mu}u_{\nu} \ .
\end{equation}

Consider the simple relativistic waves presented in \cite{shock} (here we generalize to $d$ space-time dimensions)
\bea
u^\mu=(\gamma(t,z),0...,0,\gamma v(t,z)),~~~~v(t,z)=f(z-\frac{v+c_s}{1+vc_s}t),~~~b = \left(\frac{1+v}{1-v}\right)^{\frac{c_s}{2}} \ ,
\eea
where $b$ is related to the temperature $T$ as in (\ref{b}), $f$ is an arbitrary function specifying the initial wave profile, and
$c_s$ is the speed of sound
\bea
c_s= \left(\frac{\partial p}{\partial \epsilon}\right)^{\frac{1}{2}} = \frac{1}{\sqrt{d-1}} \ .
\label{sspeed}
\eea
This configuration is a solution to equations (\ref{idealeq}), and it disperses due its implicit dependence on $v$, which causes any initial regular single-valued wave to become ultimately multiple-valued, corresponding to the crossing of characteristics \cite{shock}. Therefore it develops a singularity for any regular initial conditions function $f(z)$.
The inequality (\ref{PI}) is satisfied by the same argument as (\ref{holder}),
and gives no indication of the
singularity that will develop in the future.

\section{The viscous order}
In this section we will derive the hydrodynamic inequality at the first and second viscous order for relativistic fluids. The inequality that we will construct
is nontrivial
and can potentially be violated.

\subsection{first order in derivatives}

The correction to the relativistic hydrodynamic ideal stress-energy tensor (\ref{stressrel}) at the first viscous order reads
\bea
T_{\mu\nu}^{(1)} = -2 \eta \sigma_{\mu \nu} - \zeta \partial_{\alpha}u^{\alpha}(\eta_{\mu\nu} + u_{\mu}u_{\nu}) \ ,
\eea
where we work in the Landau frame $u^{\mu}T_{\mu\nu}^{(1)} = 0$.
$\sigma_{\mu \nu}$ is the shear tensor
\begin{equation}
\sigma_{\mu \nu}= \frac{1}{2}(\partial_\mu u_\nu+\partial_\nu u_\mu)+\frac{1}{2}(u_\mu u^\alpha \partial_\alpha u_\nu+u_\nu u^\alpha \partial_\alpha u_\mu)-\frac{1}{d-1}\partial_\alpha u^\alpha (\eta _{\mu \nu}+u_\mu u_\nu) \ .
\end{equation}
$\eta \geq 0$ and $\zeta\geq 0$ are the shear and bulk viscosities, respectively.
In the conformal case, $T_{\mu}^{\mu}=0$ implies that $\zeta=0$.
For the black brane gravitational background, the stress-energy tensor is
\bea
T_{\mu\nu}=\frac{1}{16 \pi b^d} (\eta_{\mu \nu}+du_\mu u_\nu)- \frac{1}{8 \pi b^{d-1}}\sigma_{\mu \nu} \ ,
\label{BB}
\eea
and
the ratio of the shear viscosity to the entropy density is $\frac{\eta}{s} = \frac{1}{4 \pi}$.
In the following we consider the conformal case.

The energy decreases due to shear effects and reads
\bea
E=\int \left(\frac{d\gamma^2-1}{16 \pi b^d}-2\eta \sigma_{00}\right) \ d^{d-1}x \ ,
\eea
where the first and second terms are the ideal and viscous contributions, respectively.
The decrease in the energy is written explicitly
\begin{align}
-\int 2\eta \sigma_{00} \  d^{d-1}x  &= -2 \int \eta \left(\partial_t \gamma + \gamma (-\gamma \partial_t + v^i \partial_i)\gamma - \frac{\gamma^2-1}{d-1} (-\partial_t \gamma +\partial_i v^i)\right) \  d^{d-1}x \ .
\label{energyd}
\end{align}
The entropy, on the other hand, does not change at this order, but is no longer conserved $\partial_{\mu}s^{\mu} \sim \frac{\eta}{T} \sigma_{\mu\nu}\sigma^{\mu\nu}$.
On the gravitational side, the horizon area of the deformed black brane is not corrected at the first viscous order \cite{Bhattacharyya:2008mz}.

Thus, at the first viscous order the inequality reads
\begin{align}
& \nonumber \int \frac{d\gamma^2-1}{16 \pi b^d} \  d^{d-1}x
-2 \int \eta \left(\partial_t \gamma + \gamma (-\gamma \partial_t + v^i \partial_i)\gamma - \frac{\gamma^2-1}{d-1} (-\partial_t \gamma +\partial_i v^i)\right) \  d^{d-1}x
\\
&\geq
\frac{d-1}{16 \pi}(\int \frac{\gamma}{b^{d-1}} \  d^{d-1}x  )^\frac{d}{d-1} \ ,
\label{1st}
\end{align}
 and may be violated.
Shock waves regularized by the viscosity term of the type discussed in \cite{Fouxon:2008ik}, may be possible candidates for violating the inequality
(\ref{1st}).

Note, that a fluid flow configuration that violates  (\ref{1st}) will also imply by the fluid/gravity correspondence that the deformed black brane background at first
order in the derivatives will exhibit a naked singularity. This implies on the gravity side a need for the inclusion of the higher order derivative terms, possibly an infinite number of them.

\subsection{Bjorken flow}

As a simple example one can check that the Bjorken flow (for its description see e.g. \cite{Booth:2009ct})  satisfies the inequality at the viscous order.
The Bjorken flow is a one-dimensional expansion of plasma with boost invariance symmetry along the expansion axis.
It is convenient to parametrize the flow by the proper time $\tau$ and the rapidity $y$ related to the lab frame coordinates
by
\bea
x^0 = \tau Cosh [y],~~~~~~x^1 = \tau Sinh [y] \ .
\eea

In equilibrium the energy density and the entropy density are $\epsilon = \epsilon_0 T^d$ and $s = \frac{d}{d-1}\epsilon_0 T^{d-1}$.
The inequality can be read from (\ref{NC}) with $\sigma = d-1$ and $a=\frac{\epsilon_0}{d-1}$. The solution of the hydrodynamic equations
is obtained as an expansion in $\tau^{-\alpha}, \alpha = \frac{d-2}{d-1}$. It is straightforward to check that the inequality is
satisfied at the subleading order in this expansion, which captures the first order viscous corrections.

\subsection{Second order in derivatives}

Consider next the inequality at the second viscous order.
It is straightforward to consider the most general stress-energy tensor at this order. However
for simplicity we consider the particular form of the stress-energy tensor
corresponding to the deformed black brane gravitational background, that is a conformal hydrodynamics with a particular set of transport
coefficients.
The energy reads

\begin{eqnarray}
E = \int \frac{1}{16 \pi b^d}(d \gamma^2-1-2 b \sigma_{00}-2 b \tau_\omega (u^\lambda D_\lambda \sigma_{00}+2\omega_0^\lambda \sigma_{\lambda 0})\nonumber\\
+2 b^2 (u^\lambda D_\lambda \sigma_{00}+
\sigma _0^\lambda \sigma_{\lambda 0}-
(\gamma^2-1))\frac{\sigma_{\alpha \beta}\sigma^{\alpha \beta}}{d-1}))\  d^{d-1}x \ .
\label{M2}
\end{eqnarray}
$\tau_\omega$ is a new transport coefficient that appears at this order. $\omega_{\mu \nu}$ in the vorticity tensor
\bea
\omega_{\mu \nu}= \frac{1}{2}(\partial_\mu u_\nu-\partial_\nu u_\mu)+\frac{1}{2}(u_\mu u^\alpha \partial_\alpha u_\nu-u_\nu u^\alpha \partial_\alpha u_\mu) \ ,
\eea
and $D_{\mu}$ is the Weyl-covariant derivative that acts on the velocity vector as
\bea
D_\mu u^\nu=\sigma^\nu_\mu+\omega^\nu_\mu \ .
\eea

Then the entropy current at this order reads \cite{Bhattacharyya:2008mz}
\begin{align}
\nonumber S = & \int \frac{1}{4 b^{d-1}}[\gamma+b^2 \gamma(A_1\sigma_{\alpha \beta}\sigma^{\alpha \beta}+A_2 \omega_{\alpha \beta} \omega^{\alpha \beta}+A_3 R)\\
& +b^2(B_1 D_\lambda \sigma^{0 \lambda}+B_2 D_\lambda \omega^{0 \lambda})]\  d^{d-1}x \ ,
\label{entropyc}
\end{align}
where
\bea
R = 2(d-1)\partial_{\mu}A^{\mu} - (d-2)(d-1)A_{\mu}A^{\mu} \ ,
\eea
with
\bea
A_{\mu} = u^{\alpha}\partial_{\alpha}u_{\mu} - \frac{\partial_{\alpha}u^{\alpha}}{d-1}u_{\mu} \ .
\eea
The coefficients $A's$ and $B's$ are defined in \cite{Bhattacharyya:2008mz} (Eq. (2.8) and Eq. (5.16)).

The inequality  at second order in derivatives is
\begin{align}
\nonumber &\int\frac{1}{16 \pi b^d}(d \gamma^2-1-2 b \sigma_{00}-2 b \tau_\omega (u^\lambda D_\lambda \sigma_{00}+2\omega_0^\lambda \sigma_{\lambda 0}) \\
\nonumber &+ 2 b^2 (u^\lambda D_\lambda \sigma_{00}+
\sigma _0^\lambda \sigma_{\lambda 0}-
(\gamma^2-1))\frac{\sigma_{\alpha \beta}\sigma^{\alpha \beta}}{d-1}))\  d^{d-1}x  \geq \\
\frac{d-1}{16 \pi}
\nonumber &(\int \frac{1}{b^{d-1}}[\gamma+b^2 \gamma(A_1\sigma_{\alpha \beta}\sigma^{\alpha \beta}+A_2 \omega_{\alpha \beta} \omega^{\alpha \beta}+A_3 R) \\
&+ b^2(B_1 D_\lambda \sigma^{0 \lambda}+B_2 D_\lambda \omega^{0 \lambda})]\  d^{d-1}x )^{\frac{d}{d-1}} \ .
\label{second}
\end{align}

\section{Nonrelativistic flows}

In this section we will consider the hydrodynamic inequality for nonrelativistic flows.
We will show that the inequality is always satisfied by nonrelativistic incompressible fluids, but
not when considering subleading corrections to the incompressible Navier-Stokes equations.

\subsection{Navier-Stokes equations}

The incompressible Navier-Stokes equations can be obtained in the nonrelativistic limit of relativistic hydrodynamics, where the fluid flow velocity
$v \ll c$.
Specifically we can consider relativistic CFT hydrodynamics with the stress-energy tensor (\ref{BB}). The speed of sound in relativistic CFT hydrodynamics
$v_s = \frac{c}{\sqrt{d-1}}$ and thus, we are considering in the nonrelativistic limit also the low mach limit of velocities much smaller than the speed of sound.
This is the reason for obtaining incompressible flows.

The conservation law equations of the stress-energy tensor (\ref{BB}), projected on $u_{\nu}$ and $P^{\sigma}_{\nu}$ read
\begin{align}
\nonumber & \partial_{\mu}u^{\mu} + (d-1) D ln T = \frac{1}{2\pi T}\sigma_{\mu\nu}\sigma^{\mu\nu} \ , \\
& a_{\sigma} + P_{\sigma}^{\mu}\partial_{\mu} ln T =  \frac{1}{2\pi T}P_{\sigma}^{\mu}(\partial_{\alpha}\sigma_{\mu}^{\alpha} - (d-1)\sigma_{\mu}^{\alpha}a_{\alpha}) \ .
\label{visceq}
\end{align}
In the nonrelativistic we expand $u^{\mu} = (1-v^2/2+..., v^i), T =  T_0(1+ P + ...)$ where we
scale $v^i \sim \varepsilon, \partial_i \sim \varepsilon, \partial_t \sim \varepsilon^2, P \sim  \varepsilon^2$, where $\varepsilon \sim 1/c \sim 1/v_s$.
The first equation in (\ref{visceq}) gives the incompressibility condition $\partial_i v^i = 0$, while the second equation
gives \cite{Fouxon:2008tb,Bhattacharyya:2008kq}
\beq
\partial_t v^i + v^j \partial_j v^i = -\partial^i P + \nu \Delta v^i \ ,
\label{NS}
\eeq
where $\nu = \frac{1}{4 \pi T_0}$.

\subsection{First order inequality}

The NS equations (\ref{NS}) are of order $\varepsilon^3$. The inequality at
the first viscous order is of order $\varepsilon^2$. Let us derive it. We use $b=b_0(1-P), b_0 = \frac{d}{4\pi T_0} = d \nu$. The energy up to order
$\varepsilon^2$ reads
\bea
 E = \frac{d-1}{16 \pi b_0^d}\int \left(1 +\frac{d}{d-1}{v^2}+d P\right) d^{d-1}x \ ,
\eea
where we put $c=1$.
The entropy up to order
$\varepsilon^2$ reads
\bea
4S=\frac{1}{b_0^{d-1}} \int \left(1+\frac{v^2}{2}+ (d-1)P \right)  d^{d-1}x \ .
\eea
It is easy to see that to this order the inequality is simply
\beq
\int v^2 d^{d-1}x  \geq 0 \ ,
\eeq
and is always satisfied.
This result can also be seen by noting that the viscous order correction to the energy (\ref{energyd}) is of order $\varepsilon^3$.
Alternatively, we might have anticipated the above result on general ground. At order $\varepsilon^2$ the only scalar quantities
that we can build are $v^2$, $P$ and $\partial_iv_i$. The pressure $P$ appears in the NS equations via a derivative, thus, we can expect that it will
not appear by itself, while $\partial_iv_i=0$.

A non-violation of the inequality is only a necessary condition for not having finite-time blowups, still is it curious
that it holds so simply for all incompressible flows.
On the gravity side, we may expect that the gravitational backgrounds dual to the incompressible Navier-Stokes equations will not develop
a naked singularity.
Note also in comparison, that for the relativistic flows which are compressible the inequality is not obviously satisfied.
This presumably suggests that singularities are more likely to be expected
in compressible fluid motions.

Finite-time blowups of three-dimensional nonrelativistic incompressible flows have been demonstrated with pure imaginary vector fields
in \cite{sinai}. It is amusing that in these cases the Penrose inequality is violated, unless we consider the absolute value
of the flow field.

\subsection{Second order viscous flows}

Let us consider the first subleading correction in the $1/c$ expansion to the inequality of the incompressible case,
which is now at order $\varepsilon^4$.
The fluid flows are now compressible.
For simplicity we will analyze the black brane hydrodynamics and use (\ref{second}).
The energy reads
\bea
 E=  \frac{d-1}{16 \pi b_0^d}\int\left(1 +\frac{d v^2}{d-1}+d P + \frac{d v^4}{d-1} + \frac{d^2 P v^2}{d-1} + \frac{(d+1) b_0}{(d-1)^2} v^2 \partial_i v^i \right)
 d^{d-1}x \ .
\label{E2}
\eea
Note that at this order we cannot use the incompressibility condition and the last term does not vanish.
The new transport coefficient $\tau_{\omega}$ does not appear at this order.

The entropy in the zeroth order of derivatives is
\bea
S =\frac{1}{4 b_0^{d-1}} \int \left(1+\frac{v^2}{2}+ (d-1)P +\frac{3 v^4}{8}+  \frac{(d-1)Pv^2}{2} \right) d^{d-1}x \ .
\label{S1}
\eea
The entropy is corrected at second order in derivatives (\ref{entropyc})
\bea
\int \frac{1}{4 b_0^{d-3}} \left(A_1\sigma_{ij}\sigma^{ij}+A_2 \omega_{ij} \omega^{ij}+A_3 R +
B_1 D_i \sigma^{0i}+B_2 D_i \omega^{0i} \right)  d^{d-1}x \ .
\label{S2}
\label{secondorder}
\eea
The inequality reads
\begin{align}
& \nonumber \int \frac{5v^4}{8}+\frac{dPv^2}{2}+\frac{b_0(d+1)}{d}v^2 \partial_i v^i \ d^{d-1}x \geq \\
& b_0^2 \int (\frac{A_1+A_2}{2})\partial_i v_j \partial^i v^j+(\frac{A_1-A_2}{2})\partial_i v_j \partial^j v^i-\frac{A_1}{2(d-1)}(\partial_i v^i)^2 d^{d-1}x \ ,
\end{align}
and is not obviously satisfied.

\section{Discussion}

In this paper we derived an inequality, analogous  to the Penrose inequality,
for relativistic and nonrelativistic fluid flows in a general dimension.
The inequality provides means to study singular flows. As in gravity, a non-violation of the inequality
is only a necessary condition for the absence of singularities. Indeed,
we saw that the inequality is always satisfied at the ideal fluid order, and in particular
is not violated by relativistic shock waves.

At the leading viscous order, the inequality may be violated by relativistic
fluid flows, A construction of an example that violates the inequality or a proof that it is always satisfied is still an open problem.
We saw that the inequality at this order is always satisfied by nonrelativistic incompressible flows.
Hence, on the gravity side, we may expect that the gravitational backgrounds dual to the incompressible Navier-Stokes equations will not develop
a naked singularity.
The inequality may be violated at the next to leading viscous order by both relativistic and nonrelativistic compressible flows.
This presumably suggests that singularities are more likely to be expected
in compressible fluid motions.

The derivation of the inequality is rather general and can be applied to any closed dynamical system that relaxes to equilibrium at late times.
In the context of fluid dynamics it can be easily applied also to charged hydrodynamics.
An interesting arena for a future study of the inequality is time dependent nonequilibrium relativistic dynamics.
In this case the inequality may depend crucially on the definition of the entropy, which is not uniquely defined away from equilibrium (see e.g. \cite{Romatschke:2009kr}).
Finally, there are other interesting cases where one can consider the Penrose inequality, e.g. black hole horizons with more complicated
topology.

\section*{Acknowledgements}

Y.O. would like to thank I. Bakas
for a discussion that initiated this work and N. Itzhaki for comments on the manuscript.
The work is supported in part by the Israeli
Science Foundation center of excellence, by the Deutsch-Israelische
Projektkooperation (DIP), by the US-Israel Binational Science
Foundation (BSF), and by the German-Israeli Foundation (GIF).

\end{document}